\documentclass[twocolumn,preprintnumbers,endnote,nofootinbib,prl,letterpaper]{revtex4}
\usepackage{graphicx}
\usepackage{amsmath}

\usepackage[hypertex]{hyperref}
\newcommand{\vev}[1]{\langle {#1} \rangle}
\newcommand{\lsim}{\lesssim}
\newcommand{\gsim}{\gtrsim}

\newcommand{\ord}[1]{\mathcal{O}{(#1)}}
\newcommand{\beq}{\begin{equation}}
\newcommand{\eeq}{\end{equation}}
\newcommand{\eps}{\varepsilon}

\newcommand{\trh}{T_{\rm RH}}

\newcommand{\cond}{\vev{{\bar \Psi_L} \psi_R}}
\newcommand{\condpq}{\vev{{\bar X_L} \chi_R}}
\newcommand{\eq}[1]{Eq.~(\ref{#1})}

\begin{document}

\pagestyle{plain}

\title{Technicolor Assisted Leptogenesis with an Ultra-Heavy Higgs Doublet}

\author{Hooman Davoudiasl
\footnote{email: hooman@bnl.gov}
}

\author{Ian Lewis
\footnote{email: ilewis@bnl.gov}
}

\affiliation{Department of Physics, Brookhaven National
Laboratory, Upton, NY 11973, USA}


\begin{abstract}

If fermion condensation is a main source of electroweak symmetry
breaking, an ultra-heavy Higgs doublet of mass $\sim 10^8$~GeV can
yield naturally small Dirac neutrino masses.  We show that such a
scenario can lead to a new leptogenesis mechanism based on the
decays of the ultra-heavy Higgs. Given its very large mass, the
requisite Higgs doublet can be considered an elementary particle and
would point to a cutoff scale $\sim 10^{10}$~GeV. We outline how our scenario
can also naturally lead to composite asymmetric dark matter.  Some
potential signals of this scenario are discussed.  

\end{abstract}
\maketitle

The mechanism for electroweak symmetry breaking (EWSB)  is a question of
great fundamental importance and remains a mystery.  While the Standard Model (SM) picture
based on a single Higgs doublet can accommodate all the relevant phenomenology, it could very well
be the case that Nature realizes EWSB in a more complicated way.  For example,
multiple sectors could contribute different pieces of the observed effects at low energy, with some mainly providing
$W^\pm$ and $Z$ boson masses, while others are responsible for the masses of the fermions.  In fact, some theoretical
considerations lead to such a scenario.  In particular, the apparent hierarchy between the weak scale
$\sim 100$~GeV and other potentially large scales of physics motivates one to consider a dynamical mechanism based
on condensation of fermion pairs with quantum numbers of the SM Higgs, such as those in
technicolor models \cite{Weinberg:1975gm,Susskind:1978ms}.\footnote{Another
alternative being the well-known possibility of weak scale supersymmetry which is
being tested by the LHC experiments.}  However, while a dynamical mechanism can naturally endow $W^\pm$ and $Z$
with their observed masses $m_W \sim m_Z \sim 100$~GeV,
generation of fermion masses is a challenge in this framework \cite{Dimopoulos:1979es,Eichten:1979ah,HS}.

The above considerations have provided motivation for a hybrid proposal,
namely the bosonic technicolor scenario \cite{Simmons:1988pu,Samuel:1990dq,Dine:1990jd},
where fermions obtain their masses
through Yukawa couplings with a Higgs doublet $\Phi$, as in the SM.  To see how this works in a bit more detail, let us assume
that $\Psi_L$ and $\psi_R$ are a left-handed $SU(2)_L$ doublet and a right-handed singlet, respectively, endowed
with the appropriate $U(1)_Y$ hypercharge quantum numbers to couple to $\Phi$. The Higgs potential
is then given by
\beq
V_\Phi = m_\Phi^2 \Phi^\dagger \Phi -\lambda_\psi \Phi {\bar \Psi_L} \psi_R
- \lambda_f \Phi {\bar F_L} f_R + \ldots\,,
\label{VPhi}
\eeq
where $F_L$ and $f_R$ are SM weak doublet and singlet fermions, respectively.
Upon EWSB through $\cond \neq 0$, quite generally
a vacuum expectation value $\vev{\Phi}\neq 0$ is induced for $\Phi$, given by
\beq
\vev{\Phi} = \lambda_\psi \frac{\cond}{m_\Phi^2}\,.
\label{Phivev}
\eeq
Now we have two sources of electroweak symmetry breaking: $\vev{\Phi}$ and $\cond \approx 4\pi f^3_{TC}$, where $f_{TC}$ is the techni-pion decay constant and $\vev{\Phi}^2+f^2_{TC}\approx(246~{\rm GeV})^2$.  For reasonable values of Yukawa couplings, say $\lambda_t = 2$ and $\lambda_\psi =1$, we see that the the Higgs doublet responsible for the top mass $m_t\simeq 172$~GeV can easily have a mass of a few hundred GeV to a TeV.   However, for somewhat heavier Higgs fields $\vev{\Phi}\ll m_W$
and one does not need very small Yukawa couplings to obtain the lighter fermions masses.  To avoid
reintroducing the hierarchy problem through ultraviolet quadratic quantum corrections,
this Higgs must be assumed to be composite, or else protected by a symmetry, such as supersymmetry \cite{Dine:1990jd}.
Here, we mainly assume the former possibility, but the nature of this doublet does not enter
our discussion in a crucial way.  If the Higgs field in bosonic technicolor models is to be composite,
we may expect $m_\Phi \lsim 1$~TeV and some small Yukawa couplings become necessary.

Here, we make the simple observation that the extreme smallness of neutrino masses, 
compared to other mass scales of the SM, motivates one
to treat them somewhat differently.  That is, if $m_\Phi$ is set by compositeness for all Higgs fields,
neutrino masses require very suppressed Yukawa couplings $\lambda_\nu$.
Instead, we will consider a Higgs doublet $H$ that, like other SM fields, is an elementary degree
of freedom and interacts with neutrinos through $\ord{1}$ Yukawa couplings.  This elementary Higgs
particle is then subject to large quadratic quantum corrections to its mass and is generally expected to
be very heavy.  For $m_\nu \sim 0.1$~eV, and assuming $\lambda_\nu\sim 1$, we need $\vev{H} \sim 0.1$~eV.
As before, we can have interactions of the form
\begin{eqnarray}
V_H &=& m_H^2\, H^\dagger H -\lambda_{\chi}\, H {\bar X_L}\, \chi_R- \lambda_\nu\, H^* {\bar L}\, \nu_R +  \ldots\,,
\label{VH}
\end{eqnarray}
where $X_L$ and $\chi_R$ are techni-fermions coupled to $H$, in analogy to $\Psi_L$ and $\psi_R$ coupled
to $\Phi$ in \eq{VPhi}, $L$ is a lepton doublet in the SM, and
$\nu_R$ is a singlet right-handed neutrino.  We will
assume $\lambda_\chi \sim \lambda_\nu \sim 1$.  Let us take $\condpq \sim (100~{\rm GeV})^3$.
\eq{Phivev}, applied to $H$, then yields $m_H\sim 10^8$~GeV.  We see that the requisite
mass for $H$ is quite large.  However, as mentioned before, this is a typical expectation for an elementary
Higgs, which is the origin of the hierarchy problem in the SM!  Here, assuming a typical loop-suppression,
we may infer a cutoff scale of order $\Lambda \sim 10^{10}$~GeV, relevant for the SM sector.  
A similar-in-spirit but distinct scenario in which neutrinos acquire small Dirac 
masses via heavy messenger couplings has previously been explored~\cite{hep-ph/0610275}.

From the above discussion, we see that with fermion condensation as a main source of EWSB,
an ultra-heavy Higgs doublet can provide a seesaw mechanism
for neutrino masses, in a natural way.  However, one may worry
that neutrino couplings to a TeV scale composite doublet $\Phi$, required to give masses to other SM fermions,
can spoil this picture.  Here, we simply assume that
such couplings are forbidden by a $U(1)$ Peccei-Quinn (PQ) 
symmetry \cite{PQ} under which $H$, $X_L$, $\chi_R$, and $\nu_R$ are charged.
This implies that the strongly interacting sector responsible for EWSB via fermion condensation includes other
techni-fermions, hereafter denoted by $\Psi_L$ and $\psi_R$,
that are not charged under the PQ symmetry.  These fermions can then provide masses
to the rest of the SM leptons and quarks through couplings with a composite doublet $\Phi$, as in \eq{VPhi}.  To avoid problems from an unwanted light axion we will assume that the PQ symmetry is 
not spontaneously broken at high scales.  We note that generation of mass for fermions other than neutrinos 
could be realized in other ways and the assumption of a weak scale $\Phi$ is not a necessary ingredient of our scenario; 
we only demand techni-fermion condensation and neutrino mass generation via ultra-heavy Higgs interactions.

The induced technifermion mass from Eq.~(\ref{VH}) is of the order of the neutrino mass.  Such small masses would lead to unacceptably light technipions and a light axion associated with the $U(1)_{\rm PQ}$.  However, we expect electroweak corrections to raise the technipion mass to order of electroweak scale~\cite{Chadha:1981yt}.  Also, the technipion and axion masses may be raised by physics above the TeV scale that explicitly breaks the chiral symmetry of the technifermion sector, such as in extended technicolor models~\cite{Eichten:1979ah}.  We note that the tree-level SM flavor changing neutral currents that plague extended technicolor models can be avoided, since the new chiral symmetry breaking interactions are only required  to act on technifermions (given that here the charged leptons and quarks get their masses form the TeV-scale doublet).  Hence, we expect the resulting technipions and axion to be part of electroweak-scale techni-hadron spectrum. For technicolor models with technifermions charged under $SU(3)_C$ strong interactions, Ref.~\cite{Chivukula:2011ue} used current LHC diphoton and ditau Higgs constraints to exclude technipion masses from $110$~GeV to around twice the top quark mass for a variety of models.  However, since our proposal does not contain color-charged technifermions, we expect single production via gluon fusion to be suppressed and these bounds to be considerably weaker.  Also, LEP bounds on the anomalous couplings of technipions to neutral gauge bosons can be quite constraining for technipions with masses below $\sim160$~GeV~\cite{Lynch:2000hi}. These bounds are quite model dependent and, for masses above $160$~GeV, almost non-existent, since the LEP energy reach was saturated.

As mentioned before, the generation of neutrino masses $m_\nu$ in the picture presented above is analogous to
the usual seesaw mechanism, except that the smallness of $m_\nu$ is due to
the ultra-heavy Higgs instead of the ultra-heavy right-handed
neutrino in conventional models \cite{seesaw}.  We will show that this analogy can be extended to
the possibility of leptogenesis, where it is
typically assumed that out-of-equilibrium decays of heavy right-handed neutrinos
lead to the generation of a $B-L$ charge~\cite{RIFP-641} that electroweak
sphaleron processes turn into a baryon asymmetry~\cite{IC/85/8,FERMILAB-PUB-87-034-T,Harvey:1990qw}.
Here, we will see that a similar mechanism can realize a
new kind of Dirac leptogenesis \cite{Dick:1999je,Murayama:2002je},
where the out-of-equilibrium decays of $H$ lead to $\delta (B-L)\neq 0$.

Given that the couplings $\lambda_{\chi, \nu}$ in
\eq{VH} are generally complex, the decays $H \to X_L {\bar \chi_R}$ and $H \to
{\bar L} \nu_R$ are CP violating and would lead to the generation of
an asymmetry in the fermions numbers.  In particular, if these
decays are out-of-equilibrium, a $\delta (B-L) \neq 0$ asymmetry
produced by $H$ decays will be processed into
$\delta B\neq 0$ and $\delta L \neq 0$, as long as sphalerons are in
thermal equilibrium, requiring a reheat temperature $T_{\rm RH}
\gsim 100$~GeV.  Nonetheless, complex couplings are not sufficient
for the generation of the asymmetry through CP violation;  this
also requires interference between processes at leading and
sub-leading orders that involve physical phases.  This can be
arranged if there is another Higgs particle that can contribute the necessary phase.  Note that in the absence of a Majorana mass, the one-loop vertex corrections do not contribute to the decay process of interest.  Hence, in order to
provide a mechanism for leptogenesis, we enlarge the content of the
model and assume that there are two elementary and ultra-heavy Higgs
doublets, $H_1$ and $H_2$, leading to a simple generalization of
\eq{VH}
\begin{eqnarray}
V_H& =& m_a^2\, H_a^\dagger H_a -\lambda_{\chi^D}^a\, H_a {\bar X_L}\, \chi^D_R -\lambda_{\chi^U}^a\, H_a^*{\bar X_L}\,\chi^U_R \nonumber\\
&&-\lambda_\nu^a\, H_a ^* {\bar L}\, \nu_R +
\ldots\,, \label{VHa}
\end{eqnarray}
with $a=1,2$, $m_a$ the mass of $H_a$ and we have explicitly shown the couplings of up- and down-type $\chi$ techni-quarks.

We will assume that $m_2>m_1$ and that the initial population of particles is dominated by the
symmetric production of $H_1 H_1^*$.  Hence, the effects of $H_2$ are only important through
their virtual contributions to $H_1$ decays.
To prevent the asymmetries from getting washed out, we must ensure that inverse decay
processes are decoupled from the thermal bath after $H_1$ decays have taken place.
This amounts to decoupling processes of the sort ${\bar X_L} \chi_R \to L {\bar \nu_R}$.  At
temperatures $T < m_1$, such processes are mediated by a dimension-6 operator suppressed
by $m_1^2$ and we must demand that their rate is smaller than the Hubble rate $H(\trh) = 1.7 g_*^{1/2} \trh^2/M_P$
at $T=\trh$, where $g_*$ is the number of relativistic degrees of freedom and $M_P = 1.2 \times 10^{19}$~GeV.  This
implies that $\trh$ should satisfy
\beq
\trh \lsim g_*^{1/6} \left(\frac{m_1}{M_P}\right)^{1/3} m_1.
\label{TRH}
\eeq
We then see that for $g_* \sim 100$ and $m_1 \sim 10^8$~GeV, we get $\trh \lsim 5\times 10^4$~GeV, which
is well above the electroweak phase transition temperature, but well-below $m_a$.  Thus, one must assume that
some non-thermal process, such as inflation, gives rise to a population of $H_1 H_1^*$ particles and a relatively
low-reheat initial plasma (such requirements are shared by a variety of other models; see for example 
Ref.~\cite{hylo}).  The details of the non-thermal process are not very crucial, as long as the above general
features can be obtained from it.  

For a simple estimate, let us assume a modulus field $\rho$ that couples 
universally with gravitational strength;  for example, it couples to the heavy Higgs fields through 
$(\rho/M_P) (\partial^\mu H^\dagger \partial_\mu H)$.  The width of $\rho$ is roughly estimated by 
\beq
\Gamma_\rho \sim g_*\frac{m_\rho^3}{16 \pi M_P^2}.
\label{Gamrho}
\eeq
We assume that the Universe was at some early stage in a matter dominated era due to the oscillations of $\rho$. 
These oscillations get damped by the decay of $\rho$, leading to a radiation dominated era 
at a reheat temperature estimated by 
\beq
\trh \sim (g_*^{1/2} m_\rho^3/M_P)^{1/2}.
\label{TRHEST}
\eeq
Upon the decay of $\rho$ and the subsequent prompt decays of the heavy Higgs fields, 
the SM and the techni sectors come to thermal equilibrium, at $T=\trh$, via their gauge interactions.  
Requiring $\trh \lsim 5\times 10^4$~GeV yields $m_\rho \lsim 2\times 10^9$~GeV, which 
easily allows for $\rho$ to decay into $H_1$ fields of mass $\sim 10^8$~GeV.  

\begin{figure}[tb]
\includegraphics[width=0.5\textwidth,clip]{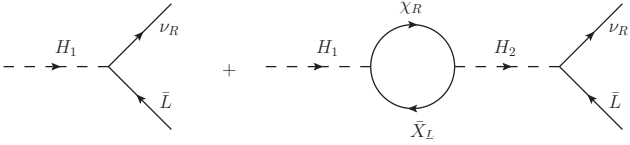}
\caption{Representative Feynman diagrams that contribute to to the non-zero value of $\eps$.}
\label{fig:1-loop}
\end{figure}
We will parametrize the asymmetry generated in the $H_1$ decays by
\beq
\eps\equiv \frac{\Gamma(H_1\to \bar L \nu_R)- \Gamma(H_1^*\to L \bar \nu_R)}{2 \Gamma(H_1)}\,,
\label{eps}
\eeq
where $\Gamma(H_1)$ is the total width of $H_1$.  For $\lambda_\chi \sim \lambda_\nu$,
we expect that $\eps \sim 1/(16 \pi^2)$, given by the interference between the tree-level
and the 1-loop amplitude for $H_1$ decay into the leptons.  For example, assuming that
diagonal couplings of $H_a$ to lepton flavors are dominant, $\eps$ is mostly
given by the contribution of diagrams of the type in Fig.\ref{fig:1-loop}.  With the techni-fermions 
in the fundamental representation of a $SU(N_{TC})$ technicolor gauge group, we find
\beq
\eps \simeq \frac{N_{TC}}{8\pi}\frac{m^2_1}{m_2^2-m_1^2}\frac{\sum_{i}{\rm Im} \left[\left({\lambda^{1*}_{\chi^D}}\lambda^2_{\chi^D}+\lambda^1_{\chi^U}\lambda^{2*}_{\chi^U}\right)\lambda^1_{\nu_i}{\lambda^{2*}_{\nu_i}}\right]}{N_{TC}(|\lambda^1_{\chi^D}|^2+|\lambda^1_{\chi^U}|^2)+\sum_{i}|\lambda^1_{\nu_i}|^2}.\,
\label{epsvalue}
\eeq
As expected, $\eps$ is of order $10^{-2}$ for $m_2 \approx 2\,m_1$ and order one couplings of $H_1$ to
leptons and techni-fermions in \eq{VHa}.

Let us now estimate the size of baryon asymmetry of the
universe (BAU) in our scenario.  After a period of inflation,
we assume that the universe gets reheated to $T_{RH}$,
through the decay of the inflaton into $H_1$ and the massless
degrees of freedom in the theory.  The prompt decays of the $H_1$
population contribute to the reheating.  However, since we would like
to maintain a low reheat temperature, that is $T_{RH}\ll m_1$, we
must require the ratio
\beq
r = \frac{n_1 m_1}{g_* T_{RH}^4}
\label{r}
\eeq
of the energy densities in $H_1$ and
radiation to be smaller than unity; here $n_1$ is the $H_1$ number
density.  We can then estimate the abundance
of $H_1$ by
\beq
Y_1 = (T_{RH}/m_1) \, r.
\label{Y1}
\eeq

As usual, we will give the BAU in terms of the ratio
\beq
\eta = \frac{n_B}{s}\,,
\label{eta}
\eeq
where $n_B$ is the baryon number density and
$s \simeq g_* T^3$ is the entropy density.  Cosmological observations have yielded 
$\eta\simeq 9 \times 10^{-11}$ \cite{PDG}.  The asymmetry
$\eps$ generated in $H_1$ decays will get processed by the various
interactions that are in thermal equilibrium in the  plasma.  In particular,
electroweak sphaleron processes will distribute an initial asymmetry
in $B-L$ (which does not get violated by any of the thermal interactions assumed here) 
and provide various other asymmetries.  
We will outline the derivation of general
formulas for such asymmetries that are relevant in our framework in the Appendix.
However, let us assume a minimal setup with
one generation of $(X_L, \chi_R)$ and $(\Psi_L, \psi_R)$
techni-fermions each, $N_\chi = N_\psi=1$,
charged under a technicolor group $SU(2)$ ({\it i.e.} $N_{TC}=2$), and only one light
Higgs doublet $\Phi$ near
the weak scale.  One can then show from the results in the Appendix that
\beq
B= \frac{13}{67} (B-L).
\label{B}
\eeq
In the above equation, $B-L$ is given by
the amount of lepton asymmetry produced in the $H_1$ decays.  It is also assumed
that at $T_{RH}$ the weak scale Higgs $\Phi$ behaves as an elementary particle 
(it is not resolved into its constituents).
For this minimal setup, we then get an estimate for $\eta$ given by
\beq
\eta \sim \frac{13}{67} \, \eps\, Y_1.
\label{eta_minimal}
\eeq
Assuming $r \sim 0.1$, $T_{RH}\sim 10^4$~GeV, $m_1 \sim 10^8$~GeV, $m_2\approx 2\,m_1$ and
adopting $\ord{1}$ couplings for $H_1$, we find
$\eta \sim 10^{-8}$ which is about two orders of magnitude larger than the observed value.
Hence, our leptogenesis model can easily account for the BAU, say,
for somewhat smaller values of couplings or slightly larger values of $m_2$.

With the minimal parameters used
for \eq{B} and the results presented in the Appendix, we also find
\beq
B_\psi = \frac{13}{201}(B-L) ,
\label{Bpsi}
\eeq
where $B_\psi$ refers to the total techni-baryon number from a single generation of
$(\Psi_L, \psi_R)$ fermions.  Let us assume that these fermions form the lightest techni-baryon
$S=\Psi^u \psi^d$ with zero electric charge.  If we also assume that all
the interactions that would violate $B_\psi$ are sufficiently suppressed,
in analogy with the SM proton decay operators, the associated $S$-baryons are
cosmologically stable.  The above result (\ref{Bpsi})
then suggests that such a particle made of $(\Psi_L, \psi_R)$
could be a good dark matter (DM) candidate.

Since the energy density in DM is about 5 times larger
than that in ordinary baryons, \eq{B} and \eq{Bpsi} imply that with a mass $m_S \sim 15$~GeV
$S$ could be a good DM candidate.  However, most likely, $m_S \sim 1$~TeV,
given that we expect $\cond \sim (100~{\rm GeV})^3$.  This seems to suggest that a suppression of
$\ord{10^{-2}}$ in $B_\psi$ is necessary, so that $S$ can have the required cosmological
energy density.  Remarkably, given a reasonable value for $T_c\sim 200$~GeV, the
sphalerons will typically lead to a suppression of order
$(m_S/T_c)^{3/2}e^{-m_S/T_c}\sim 10^{-2}$ \cite{Barr:1990ca,dm-supp}.  Hence,
we see that our leptogenesis  mechanism can, in principle,
naturally lead to a good asymmetric DM candidate $S$ \cite{ADM}.
In any event, the viability of the DM candidate in
our scenario depends on the details of its specific
implementation, which is outside the main scope of the current work.

It may also be possible that techni-baryon number is violated by higher dimensional operators and techni-baryons are unstable.  In such a case, the decay of the primordial techni-baryons into light SM particles will cause a large increase in the entropy of the early universe.  If this decay occurs during or after Big Bang Nucleosynthesis (BBN), the increase in entropy will strongly perturb the abundances of the light elements.  Hence, the techni-baryons must either decay before BBN, i.e., $\tau_{TB}\ll 1$~second, where $\tau_{TB}$ is the techni-baryon lifetime, or be long- lived on cosmological time scales.  Assuming that technibaryon number is violated by a dimension six operator, the first scenario leads to the condition
\beq
\frac{m_S^5}{M^4}\gtrsim 10^{-24}~{\rm GeV}.
\label{dim6.eq}
\eeq
Hence, for $m_S\sim 1$~TeV, the cutoff for the technibaryon violating process is $M\lesssim10^{10}$~GeV, close to the cutoff for the SM sector.  In the second, ``long-lived", scenario, agreement with observation require $\tau_{TB}\gtrsim 10^{26}$~sec.  Such a case would lead to the interesting possibility of decaying DM~\cite{Nardi:2008ix}. For $m_S\sim 1$~TeV, Eq.~(\ref{dim6.eq}) implies the cutoff is then $M\gtrsim 10^{16}$~GeV, near the Grand Unified scale.

Although the mechanism for neutrino mass generation 
is far out of the reach of present experimental searches, the model presented here is still falsifiable and may have some signatures at the LHC.    First, this scenario generates Dirac neutrino masses.  Hence, if neutrinos are determined to be Majorana, for example through observation of neutrinoless double $\beta$-decay~\cite{42670}, our model will be ruled out.

Since technicolor is the main source of EWSB, we would expect to see TeV scale techni-hadrons at the LHC.  In the scenario presented here technicolor was paired with a composite Higgs.  For this specific realization, Higgs like scalars may also be accessible at the LHC.  As mentioned earlier, for reasonable values of $\lambda_t$ and $\lambda_\psi$, the composite Higgs scalars have masses on the order of several hundred GeV to a TeV.  Compared to the SM, the composite Higgs has a suppressed coupling to $W/Z$; hence, traditional searches~\cite{HiggsCombo} for a high mass Higgs boson may need to be modified.  It is also possible for the Higgs like scalars to have masses near the LEP Higgs mass bound $114$~GeV~\cite{Barate:2003sz}.  In that case, using a holographic approach, it has been shown that in a bosonic 
technicolor model similar to ours, LEP data and EW precision constraints 
bound the techni-hadrons to have masses above $\sim 2$~TeV and techni-pion decay constant $f_{TC}\lesssim100$~GeV~\cite{arXiv:1003.4720}.  The Higgs like scalar may then have couplings to SM particles similar to those of the SM Higgs boson and, hence, may have similar signatures as the SM Higgs at the LHC. 
  However, we stress that this neutrino mass scenario does not rely on the mechanism through which the other fermions gain mass, i.e., it can be paired with any viable technicolor model.

We examined the possibility that dynamical electroweak symmetry breaking, as in technicolor models, 
could provide Dirac masses for neutrinos via an ultra-heavy Higgs doublet of mass $\sim 10^8$~GeV,  
with couplings of order unity.  The hierarchic mass scale of this doublet suggests 
it should be considered an elementary degree of freedom, 
far above the weak scale.  Adopting the bosonic technicolor framework 
for illustrative purposes, we showed that the CP violating 
decays of the ultra-heavy Higgs scalar can provide a novel mechanism for leptogenesis.  
Typical parameters in our setup can yield the correct cosmological baryon number.  This setup, under some conditions, 
can also lead to a viable asymmetric dark matter density made up of techni-baryons.  
Our model implies the emergence of techni-hadrons at the TeV scale.  In a bosonic technicolor framework 
one would also expect the appearance of composite Higgs-like scalars at the weak scale, but with non-Standard-Model-like   
interactions, which could be studied at the LHC.  Quite generally, 
the observation of neutrinoless double $\beta$-decay can rule out 
the scenario introduced here.


\acknowledgments
We thank David Morrissey for comments.  This work is supported in part by the US DOE Grant DE-AC02-98CH10886.
\appendix
\section{Baryon Number Calculation}
 
The asymmetry between particle and anti-particle density  is proportional to the particle's chemical potential, $\mu_i$.  Hence, only relationships between chemical potentials need to be calculated.  Here we comment on properties peculiar to our scenario.  Generic details of the calculation can be found in Ref.~\cite{Harvey:1990qw}. 

As noted previously, sphaleron processes are expected to contribute to rapid fermion number violation at temperatures $T>T_c$.  These interactions will create $N_f$ baryons and leptons, and $N_{\psi}$ and $N_{\chi}$ techni-baryons of $\psi$ and $\chi$ type, respectively.  When interactions are in thermal equilibrium the sum of the chemical potentials of the incoming particles is equal to the sum of the outgoing.  Hence, for $T>T_c$ sphaleron processes imply
\begin{eqnarray}
0&=&\frac{N_{TC}}{2}\sum_i(\mu_{\chi^U_{iL}}+\mu_{\chi^D_{iL}})
+\frac{N_{TC}N_\psi}{2}(\mu_{\psi^U_L}+\mu_{\psi^D_L})\nonumber\\
&&+N_f\,(2\,\mu_{dL}+\mu_{uL})+\sum_i\mu_{\nu_{iL}}.
\end{eqnarray}
Flavor changing Yukawa interactions equalize the chemical potentials of the $\psi^{U}$, $\psi^D$ and quark generations and we use one chemical potential for each particle type.  At the reheat temperatures we are interested in, the flavor changing interactions of neutrinos, $X_L$ and $\chi_R$  are out of equilibrium; hence, the generational chemical potentials are kept distinct.

Following the usual arguments for $B-L$ conservation, we find that $N_\chi L-N_f B_{X_L}$ and $N_\psi L-N_f B_\psi$ are also conserved, where $L$ is charged lepton and $\nu_L$ number, and $B_{X_L}$ is the $X_L$ techni-baryon number.  We expect $\chi_R$ and $\nu_R$ numbers to be separately conserved since the reheat temperature is below the energy at which interactions mixing $\chi_R$ or $\nu_R$ with other species are in thermal equilibrium.  Finally, we note that if in Eq.~(\ref{VHa}) $\lambda_{\chi^D}=\lambda^*_{\chi^U}$, then 
\beq
B-L=\frac{N_f}{N_\chi}B_{X_L}-L=\frac{N_f}{N_\psi}B_\psi-L=B_{\chi^D_R}-B_{\chi^U_R}=-L_{init},
\eeq
where $L_{init}$ is the initial lepton number injected by $H_1$ decays.  Once the algebra is accomplished, one obtains Eqs.~(\ref{B}) and (\ref{Bpsi}).

\end{document}